A Likelihoodist Trial Procedure Performs Well Under Simulation

Author: Nicholas Adams

Institution: Alfred Health, Melbourne, Australia.

Abstract

A simple and common type of medical research involves the comparison of one treatment against another. The logical aim should be both to establish which treatment is superior and the strength of evidence supporting this conclusion, a task for which null hypothesis significance testing is particularly ill-suited. This paper describes and evaluates a novel sequential inferential procedure based on the likelihood evidential paradigm with the likelihood ratio as its salient statistic. The real-world performance of the procedure as applied to the distribution of treatment effects seen in the Cochrane Database of Systematic Reviews is simulated. The misleading evidence rate was 5% and mostly this evidence was only weakly misleading. Early stopping occurred frequently and was associated with misleading evidence in only 0.4% of cases.

Introduction

Medical research commonly involves the comparison of two treatments (one of which may be a placebo), where the sampling distribution of the difference is approximately normal. The purpose of such research is to determine either the superiority or non-inferiority of one treatment versus the other. We describe a novel, sequential likelihoodist inferential procedure to achieve this, one that produces a likelihood ratio (LR) as a natural measure of the strength of evidence supporting such a determination. We discuss the inferential interpretation of such an LR and confirm its theoretical performance via simulation. Current inference from such trials is almost always performed using null hypothesis significance testing (NHST) and p-values. Unfortunately, misinterpretation with this approach is both persistent and pervasive.[1] We suggest that a directional likelihood ratio calculated using the minimum clinically significant effect size as a dividing hypothesis provides a more useful and easily understood metric.

Methods

- Pre-data, specify a minimum clinically significant effect size ($\Delta$) for the parameter of interest ($\theta$), which compares the two treatments. The direction of the effect is not pre-specified.

- Collect and sequentially analyse data – in the likelihood evidential paradigm there is no adjustment for multiple looks at the data.[2,3]

- For each data point a one-sided p-value is calculated against the dividing hypothesis $\theta=\Delta$ using the observed effect size ($\theta_{obs}$).[4] This p-value is converted into a directional likelihood ratio (LR).[5,6] For a symmetric sampling distribution:

$$LR = \frac{0.25}{p - p^2} \qquad (1)$$

By convention we use this LR when $\theta_{obs} > \Delta$, and its inverse (1/LR) otherwise. The LR then is a measure of the strength of evidence that the true effect size ($\theta_T$) is greater than $\Delta$.

- Data is collected until we obtain LR >20 or LR<0.05. This LR corresponds to an effect size of ~ 2.25 standard errors and a one-sided p-value of ~ 0.012.

- Note the limit of LR as the sample size increases:

$$\lim_{n \to \infty} (LR) = \begin{cases} \theta_T > \Delta : \infty \\ \theta_T = \Delta : undefined \\ \theta_T < \Delta : 0 \end{cases}$$

and the expectation, E(LR)=1 when $\theta_T = \Delta$. In the unlikely case that $\theta_T = \Delta$ we may not ever attain LR>20 or LR<0.05 to allow us to stop sampling, or, if the value of $\theta_T$ is close to $\Delta$ we may require an unfeasibly large sample size to do so.[3] To avoid these problems, it is necessary to specify a maximum sample size. This maximum may be determined by resource constraints or, preferably, based on precision as discussed below.

- Conversely, small sample sizes when filtered by a stopping rule such as 20<LR<0.05 will lead to over-estimation of the value of $\theta_T$.[7] For this reason, it is necessary to specify a minimum sample size, again based on precision.

- When we stop sampling, the summary statistic of the all the data observed is the final LR which will be less than 1 if $\theta_{obs}<\Delta$, and greater than 1 if $\theta_{obs}>\Delta$, providing the strength of evidence supporting the proposition that the true effect size exceeds the minimum clinically significant effect size.

The minimum clinically significant effect size ($\Delta$) is an important value which should be determined clinically rather than statistically.[8] In the current medical literature, it is common for this value to be specified in the context of the pre-study determination of sample size, but never mentioned again. In our procedure, $\Delta$ remains prominent in the final reported summary statistic of the trial.

Sequential analysis has obvious practical benefits – a research study with an unexpectedly large effect size can be stopped early, avoiding wastage of time and resources.[9] The choice of 20<LR<0.05 as a stopping rule is arbitrary and chosen to nominally align with the traditional 0.05 type 1 error threshold. As Royall has shown, the probability of misleading evidence is automatically controlled by the use of likelihood ratios.[10] Royall's work was based on the comparison of two simple point hypotheses, but we extend this to a dividing hypothesis using the generalised law of likelihood for composite hypotheses.[11,12] In our simple model, dividing the parameter space at $\Delta$ means the supremum of the likelihood function on one side of the division is the likelihood at the observed effect size and the supremum of the other side is the likelihood at Δ. Note also that the procedure functions simultaneously as both a superiority trial (if LR>20) and a non-inferiority trial (if LR<0.05).

Conventional sequential probability ratio tests are variations of classical Neyman-Pearson error control and are designed to provide a trichotimised answer to the hypothesis posed (accept, reject, or indecisive).[13] By contrast, our procedure is designed to provide strength of evidence for or against acceptance of the hypothesis ($\theta_T > \Delta$), with no explicit error control. Sequential Bayes factor testing differs by the need to specify a prior probability distribution.[9] Sequential analysis is not an essential part of our procedure. Indeed, a likelihood ratio can retrospectively be calculated from a NHST designed trial if the sampling distribution is normal. Given a standardised effect size (Z),

$$LR = \frac{0.25}{\Phi(Z - \Delta) - [\Phi(Z - \Delta)]^2} \quad (2)$$

where $\Phi$ is the Gaussian cumulative distribution function.

While it is necessary for practical reasons to specify a maximum sample size, it is also important that the sample size is large enough that the precision of estimation of the true effect size is sufficient. Traditional sample size and power calculation are not possible in our procedure because the p-value is calculated against the minimum clinically significant effect size. Instead, we calibrate the sample size using a 95% confidence interval.[14,15] We base the calibration on two logical principles. The first is that 95% confidence interval should be smaller than 2Δ to avoid the potential ambiguity of the interval including the value of Δ in both directions simultaneously. Thus, the minimum sample size (n) is:

$$n = \left\{\frac{1.96}{\Delta}\right\}^2$$

The second principle is that the 95% confidence interval need not be smaller than Δ. It is implicit in the specification of Δ that we are indifferent to whether $\theta_T=0$ or $\theta_T=\Delta$ or any value in between, and we suggest this margin of indifference should also apply to the span of a 95% confidence interval. Thus, the maximum sample size is:

$$n = \left\{\frac{2 \times 1.96}{\Delta}\right\}^2$$

The procedure is therefore to sample at least to the minimum sample size then sequentially analyse, stopping when LR>20 or LR<0.05, or when the sample size reaches this maximum. Using NHST researchers may sometimes be tempted to unreasonably inflate Δ to achieve a nominated power at a smaller sample size than would otherwise be required. In our procedure the effect of increasing Δ, while reducing the minimum and maximum sample size, also makes it harder to achieve LR>20 so as to stop sampling early. Thus, the temptation of sample size parsimony is removed because a larger value for Δ is not guaranteed to lead to a smaller final sample size.

Likelihood ratios provide a natural and intuitive measure of the strength of evidence.[16-18] Many medical trials are performed from a position of equipoise where the prior odds of the

superiority of one treatment over another are approximately even (1:1). The odds form of Bayes theorem is:

$$posterior\ odds = prior\ odds \times likelihood\ ratio$$

and hence if the prior odds are 1:1 then the posterior odds equal the likelihood ratio. Most people are familiar with betting odds and will have little trouble interpreting LR=1/10 as odds of 10:1 against. The conventional interpretation of NHST is that a significant p-value is sufficient proof that one treatment is better than another, and a non-significant p-value is not. Implicit is that a decision should be made to use the better treatment when statistical significance is obtained, and the presence or absence of clinical significance is ignored. By contrast, our LR is a continuous measure of strength of evidence favouring a clinically significant difference between the treatments. The decision to use a particular treatment then involves weighing the LR against its potential harms and costs.

Finally, from a logical standpoint, likelihood ratios are both coherent and consistent, a claim which cannot be made for 2-sided p-values nor Bayes factors.[19-21]

Results

We sought to estimate the real-world performance of our procedure by simulating inference from a sample of randomised controlled trials. The distribution of standardised effect sizes (Z-scores) in the large Cochrane Database of Randomised Controlled Trials is approximately normal (actually a mixture of normals), with a mean close to zero and a standard deviation of ~1.8.[22] These trials represent an elite subset of all medical trials likely with higher nominal power than usual.[23] For this reason, and to be conservative, we chose to simulate 1000 trials from a narrower distribution with a Z-score standard deviation of 1.0 and mean of zero. We used Microsoft Excel for Mac version 16.5, sequentially analysing using the procedure described above, and a standardised minimum clinically significant effect size of 0.5. We defined misleading evidence as being a final LR in the wrong direction, that is favouring $\theta > \Delta$ when $\theta_T < \Delta$, or vice versa. The results of the simulation are shown in Table 1. Overall, the mean sample size was n=39. This simulation demonstrates that the procedure produces misleading evidence about 5% of the time and almost always this misleading evidence is weak. At the same time, it reduces resource wastage by stopping early when the difference between $\theta_T$ and $\Delta$ is large (Figure 1). The results of the simulation of course depend upon the distribution of effect sizes and the value chosen for $\Delta$, however they demonstrate the feasibility of the procedure under fairly typical conditions.

Conclusion

Our procedure represents an alternative to NHST for simple trials involving the comparison of two groups where the sampling distribution is approximately normal. The advantages of our approach include:

- Encourages reasonable specification of the minimum clinically significant effect size

- Allows unpenalized sequential analysis

- Provides a logical rationale for the minimum and maximum sample sizes

- Produces an intuitive summary statistic that is measures the weight of evidence supporting clinical significance

The feasibility of the approach is confirmed by simulation.

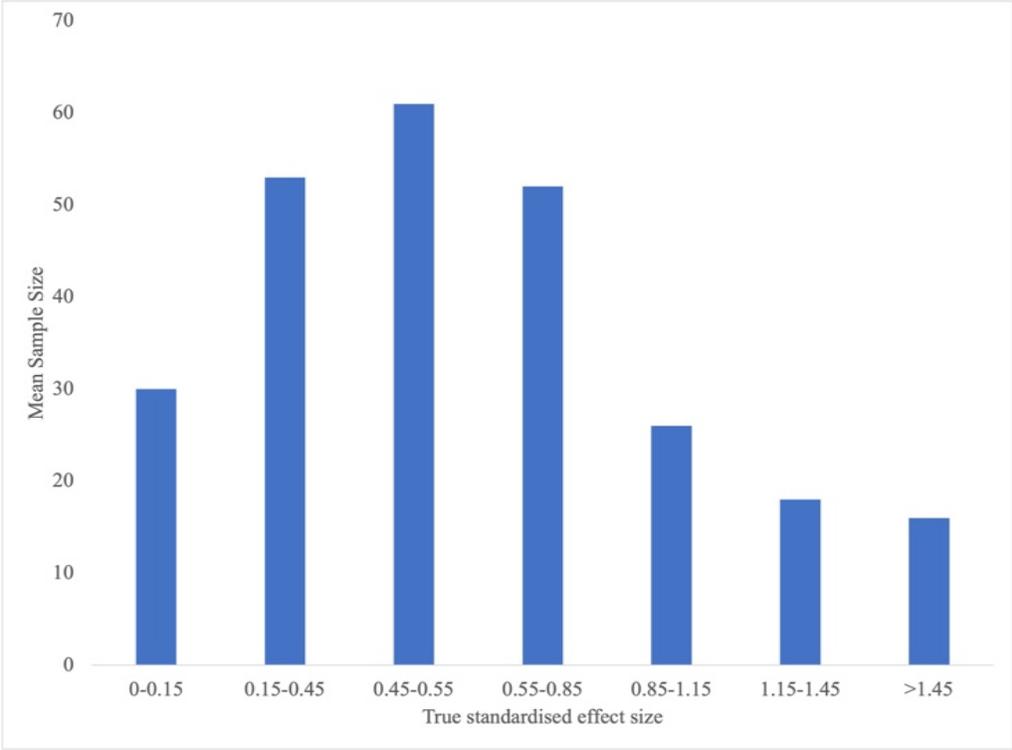

Figure 1. Mean sample size at termination of the procedure for various true effect sizes under simulation, minimum clinically significant effect size (Δ) =0.5.

| Result | Incidence (%) | Mean LR |
| --- | --- | --- |
| **Misleading evidence, stopped early (N<64)** | 0.4 | >20 |
| **Correct evidence, stopped early (N<64)** | 59 | >20 |
| **Misleading evidence, stopped at N=64** | 4.7 | 1.8 |
| **Correct evidence, stopped at N=64** | 36 | 3 |

Table 1. Results of n=1000 simulated trials. Misleading evidence is defined as a likelihood ratio (LR) favouring $\theta>\Delta$ when $\theta_T<\Delta$, and vice versa. (N=sample size)